\documentclass[floatfix,twocolumn,pre,showpacs]{revtex4}
\usepackage{graphicx}
\usepackage{subfigure}
\usepackage{amssymb,amsmath}

\begin{document}

\title{Capillary filling in patterned channels}
\author{H. Kusumaatmaja${}^{a}$} 
\author{C. M. Pooley${}^{a}$}
\author{S. Girardo${}^{b}$}
\author{D. Pisignano${}^{b}$}
\author{J. M. Yeomans${}^{a}$}
\affiliation{${}^{a}$Rudolf Peierls Centre for Theoretical Physics, 1 Keble Road, Oxford, OX1 3NP, United Kingdom}
\affiliation{${}^{b}$NNL, National Nanotechnology Laboratory of Istituto Nazionale di Fisica della Materia-CNR and Scuola Superiore ISUFI, Universit\`{a} del Salento, via Arnesano, I-73100 Lecce, Italy}
\date{\today}


\begin{abstract}
We show how the capillary filling of microchannels is affected by posts or ridges on the
sides of the channels. Ridges perpendicular to the flow direction introduce contact line
pinning which slows, or sometimes prevents, filling; whereas ridges parallel to the flow 
provide extra surface which may enhances filling. Patterning the microchannel surface 
with square posts has little effect on the ability of a channel to fill for equilibrium contact 
angle $\theta_e \lesssim 30^{\mathrm{o}}$. For $\theta_e \gtrsim 60^{\mathrm{o}}$, 
however, even a small number of posts can pin the advancing liquid front.
\end{abstract}
\pacs{47.61.-k, 47.55.nb, 68.08.Bc}
\maketitle


Recent years have seen rapid progress in the technology of fabricating channels at micron  
length scales. Such microfluidic systems are increasingly finding applications, for  
example, in diagnostic testing and DNA manipulation and as microreactors. As narrower  
channels are used, to conserve space and reagents, surface effects will have an increasing    
influence on the fluid flow. In particular, it may be possible to exploit the chemical or  
geometrical patterning of a surface to control the flow within the channels. Indeed first  
steps in this direction have already been taken: for example, Joseph {\it{et al.}} used  
superhydrophobic surfaces to introduce slip lengths of the order a few $\mu$m \cite{Joseph},
Zhao {\it{et al.}} used hydrophilic stripes to confine fluid flows \cite{Zhao}, and   
Stroock {\it{et al.}} used oriented grooves to enhance mixing \cite{Stroock}.  

In this paper we describe the way in which a fluid penetrates a microchannel patterned  
with posts or ridges. We find that the hysteretic behaviour that results from the surface  
patterning can be a substantial factor in controlling how the fluid moves down the  
channel. Ridges or posts oriented parallel to the flow speed up capillary filling while  
those perpendicular to the flow pin the interface and suppress, or sometimes prevent, 
filling. We investigate the rate at which the microchannel fills, finding that it is  
highly dependent on both the geometry and the spacing of the posts. In particular, we  
demonstrate that a patterned surface can act as a valve, allowing filling in one direction  
only. 


The classical analysis of capillary penetration is due to Washburn \cite{Washburn}.  
Consider a capillary of width $h$ with an infinite reservoir of liquid of dynamic  
viscosity $\eta$ at one end. If the walls of the capillary are hydrophilic the liquid will  
move to fill it with a mean velocity (assuming two dimensional Poiseuille flow), 
$\bar{v} = - \frac{h^2}{12 \eta} \frac{dp}{dx}$, where $\frac{dp}{dx}$ is the pressure 
gradient that sets up the flow. The driving force for the flow is provided by the decrease 
in free energy as the fluid wets the walls or, equivalently, the Laplace pressure across 
the interface. Hence $\frac{dp}{dx} = - \frac{\gamma}{Rl}$,
where $\gamma$ is the surface tension of the liquid interface, $R = h/2\cos{\theta_a}$ is its  
radius of curvature and $l$ is the length of liquid in the tube. Substituting $\frac{dp}{dx}$
into the parabolic flow profile and using $\bar{v} = dl/dt$ gives Washburn's law
$l=\left(\gamma h \cos{\theta_a}/3 \eta \right)^{1/2} \left({t+t_0}\right)^{1/2}$, 
where $t_0$ is an integration constant. Note that it is appropriate to use, not the  
static, but the advancing contact angle  $\theta_a$, as this controls the curvature of the  
interface and hence the Laplace pressure. Washburn's law assumes that there is no resistance 
to motion from any fluid already in the  
capillary.  
The derivation also assumes incompressibility, and neglects inertial effects  \cite{Quere,Succi}, gravity  \cite{Timonen} 
and any deviations from a Poiseuille flow profile at the inlet or the interface 
\cite{Succi,Levine,Szekeley,Dimitrov}.


Our aim is to explore capillary filling in the presence of topological patterning on the  
surface of the capillary. Simple arguments, based on the Gibbs' criterion \cite{Gibbs}, will suffice to  
explain when capillary filling is not possible for one dimensional surface patterning  
(ridges). However, to see how the rate of filling is affected by the surface relief, and  
to consider two dimensional surface topologies, we will need to solve the hydrodynamic 
equations of motion  of the fluids. Therefore we first describe the numerical model we 
shall use, a diffuse  interface representation of a binary fluid, and check that it reproduces 
Washburn's law for simple capillary filling. 

We consider an immiscible binary fluid, comprising components $A$ and $B$ say, described, in  
equilibrium, by the Landau free energy
\begin{equation}
\int_V dV \left(\frac{\alpha}{2}\phi^{2}+\frac{\beta}{4}\phi^{4}+\frac{\kappa}{2}(\nabla\phi)^{2} + T\rho\ln\rho \right) -  \int_S dS \left(\sigma \phi \right) \label{free}
\end{equation}
where $\rho = \rho_A+\rho_B$ is the total fluid density, and $\phi = \rho_A-\rho_B$ is the  
concentration. We chose $\alpha = -\beta$ so that the two possible bulk concentrations are $\phi =  
1.0$ and $\phi = -1.0$. $\kappa$ is related to the surface tension by $\gamma =  
\sqrt{8\kappa \alpha/9}$ and $T$ is chosen to be 1/3 to minimise the error
terms. The surface contribution allows us to tune the equilibrium contact angle 
by varying $\sigma$ \cite{Briant}.
 
The fluid motion is described by the continuity, Navier-Stokes and convection-diffusion  
equations
\begin{eqnarray}
&\partial_{t} \rho + \partial_{\beta} \rho u_{\beta} = 0, \label{continuity} \\
&\partial_{t} \rho u_{\alpha} + \partial_{\beta} \rho u_{\alpha} u_{\beta} =  
-\partial_{\beta} P_{\alpha\beta} + \partial_{\beta} S_{\alpha\beta}, \label{N-S}\\
&\partial_{t}\phi({\bf x},t) + \partial_{\alpha}\phi u_{\alpha} =   
M\partial_{\alpha}^{2}\mu({\bf x},t). \label{diffusion}
\end{eqnarray}
where $\vec u$ is the fluid velocity, $P_{\alpha \beta}$ and $S_{\alpha \beta}$ are the  
pressure and stress tensors respectively, $\mu$ is the chemical potential and $M$ is a  
mobility. We solve these equations using a lattice Boltzmann algorithm, described in  
detail in \cite{Briant}. Similar models have also been recently used in  
\cite{Succi, Timonen} to simulate capillary filling on smooth surfaces.

\begin{figure}
\includegraphics[scale = 0.71]{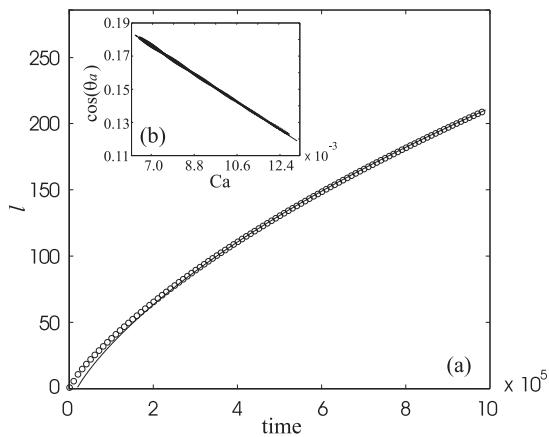}
\caption{Lattice Boltzmann simulation results for capillary filling for $h = 50$, 
$\theta_e = 60^{\mathrm{o}}$, $\gamma = 0.0188$, $\rho_{\mathrm{A}} = \rho_{\mathrm{B}} = 1.0$, 
$\eta_{\mathrm{A}} = 0.83$, $\eta_{\mathrm{B}} = 0.03$ and $M = 0.05$. (a) The length of the column of 
the filling $A$-component plotted against time. The circles are the simulation results and the 
solid line is a fit to the Washburn's law using the measured advancing contact angle
and correcting for the small viscosity of the displaced $B$-component. 
(b) The measured advancing contact angle of the liquid-liquid interface: note that $\cos{\theta_a}$ varies
linearly with capillary number \cite{Rothman}.}
\label{fig1}
\end{figure}
Numerical results showing capillary filling of a smooth channel are presented in 
Fig.~\ref{fig1}(a). The plot is for a channel of length $L=640$, infinite width and height  
$h=50$. Reservoirs ($480 \times 200$) of  components $A$ and $B$ are attached at each end  
of the capillary. The two reservoirs are connected to ensure that they have the same pressure. 
The parameters of the model (given in the caption of Fig. \ref{fig1}) are chosen so that the 
capillary and Reynolds numbers are of order $10^{-2}$.

The solid line in Fig.~\ref{fig1}(a) is a fit to the Washburn's law using the measured 
values of the contact angle and correcting for the small viscosity of the displaced $B$-component. 
The fit is excellent, except very close to the beginning of the simulation where the expected 
deviations from Washburn's law due to inertial effects and deviations from a Poiseuille flow 
profile are observed. Fig. \ref{fig1}(b) shows the measured $\cos{\theta_a}$ which 
decreases linearly as the capillary number decreases \cite{Rothman}. 

 
\begin{figure*}
\includegraphics[scale = 0.65]{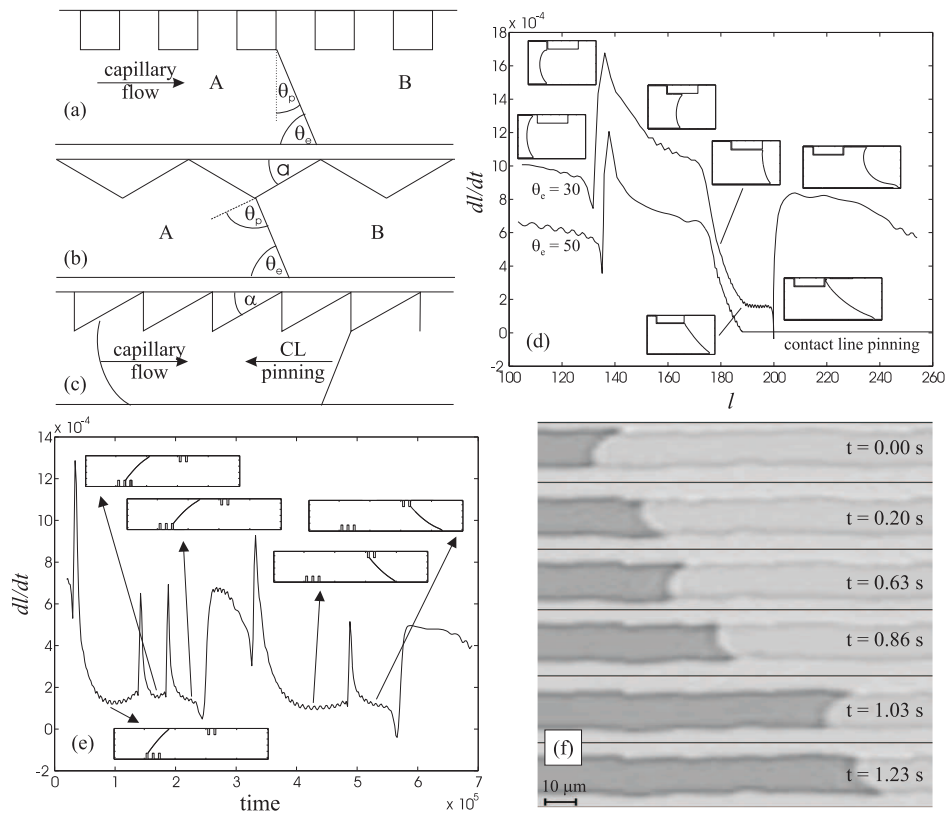}
\caption{Pinning of the contact line during capillary filling when the top wall is patterned 
with (a) squares, (b) triangles, and (c) ratchets. (d) Interface velocity $dl/dt$ when one of 
the walls is decorated with a single $40 \times 10$ post. (e) Interface velocity when 
the bottom wall is decorated with three posts and the top wall is decorated with two posts. 
The posts have height $10$ and width $5$. (f) Experimental results on capillary filling 
in patterned channels. Local pinning temporarily holds the capillary 
front on one of the two microchannel sidewalls, while the interface advances along the opposite 
sidewall.}
\label{fig3}
\end{figure*}
We now discuss capillary filling in a channel with topological patterning on the surface.  
On stepped surfaces the driving force of the Laplace pressure must be sufficient to  
overcome any hysteresis due to the  surface patterning. We shall assume that the contact  
angle of the $A$-component is $ < 90^\mathrm{o}$ i.e. this is the fluid that may fill the  
capillary, and that there is no intrinsic hysteresis.
 
We first consider two dimensions, and a channel where one of the walls is patterned  
by square posts, as shown in Fig.~\ref{fig3}(a). (The same results apply for a three  
dimensional channel with ridges which are translationally invariant perpendicular to the  
flow.) According to the Gibbs' criterion, for an interface moving very slowly, the contact  
line will remain pinned on the edge of the posts until $\theta_p$, the angle the interface  
makes with the side of the posts, reaches the equilibrium contact angle $\theta_e$. It 
follows immediately (see Fig.~\ref{fig3}(a)) that there will be no capillary filling unless
\begin{equation}
\theta_p \equiv 90^o - \theta_e < \theta_e, \label{Pinning}
\end{equation} 
that is, $\theta_e < 45^{\mathrm{o}}$. If the surface is patterned by triangular posts,  
with sides at an angle $\alpha$ to the edges of the channel as shown in Fig.~\ref{fig3}(b) ,
the condition for capillarity is less stringent, namely $\theta_e < 90^o -\alpha/2$. Hence  
the surface in Fig.~\ref{fig3}(c) will allow filling from left to right, but not from  
right to left if $45^o  < \theta_e < 90^o -\alpha /2$.

Numerical results showing how the interface behaves as it approaches a post with finite  
velocity are presented in Fig.~\ref{fig3}(d) for channel height $h=50$, post height $h_p=10$, 
post width $w_p=40$ and equilibrium contact angles $\theta_e = 30^{\mathrm{o}}$ and $50^{\mathrm{o}}$. 
Here $l$, the length of the liquid column, is measured at the centre of the capillary channel. 
As the interface approaches the post, it is slowed down by dissipation caused by the post. 
However, once the interface is able to respond to the wetting potential of  
the post, it quicky wets the leading edge. The interface slows as it moves over the top  
the post -- if the post was infinite in extent it would reach the speed appropriate to a  
channel of height $h-h_p$. Once the trailing edge of the post is reached there is a  
substantial decrease in velocity as the interface becomes pinned. For $\theta_e =  
30^{\mathrm{o}}$ the interface is able to proceed, but for $\theta_e = 50^{\mathrm{o}}$
it remains pinned, consistent with the condition in Eq.  (\ref{Pinning}). For the same value  
of $\theta_e$, the variation in velocity is more pronounced for higher posts.

Fig. \ref{fig3}(e) shows the dynamics as the interface moves along a capillary with
groups of posts on both the top and bottom surfaces of the channel. This simulation 
demonstrates several features of the motion. First, as the interface reaches the three
posts on the lower edge of the channel, it takes longer to depin from the first post it 
reaches.  This is because the free end of the interface takes time to slide along the 
top of the capillary until it reaches an angle where it can pull the pinned end away from 
the trailing edge of the post. For the second and third posts the interface is already 
aligned diagonally across the channel and does not need to move so far. Once the 
interface reaches the posts at the top of the channel it remains pinned on the first of 
these for some time until the interface again takes up a diagonal configuration, but 
now with the bottom edge leading. 

Such switching between two diagonally aligned interface configurations has been observed in experiments on capillary filling in channels with lateral root-mean-square roughness 1 $\mu$m, as shown in Fig. \ref{fig3}(f). The channels were realised by optical lithography employing plastic photomasks patterned with a laser printer. Polymer microchannels were then obtained by subsequent replica molding with polydimethilsiloxane. A 2 $\mu$l drop of water was released at the entrance of the microchannels and the penetrating interface was imaged by a 320$\times$240 pixels, 30 fps camera.
Experiments have also shown the interface waiting 
longer at the beginning of a line of posts engineered to lie on one surface of a microchannel. 


\begin{figure}
\includegraphics[scale = 0.45]{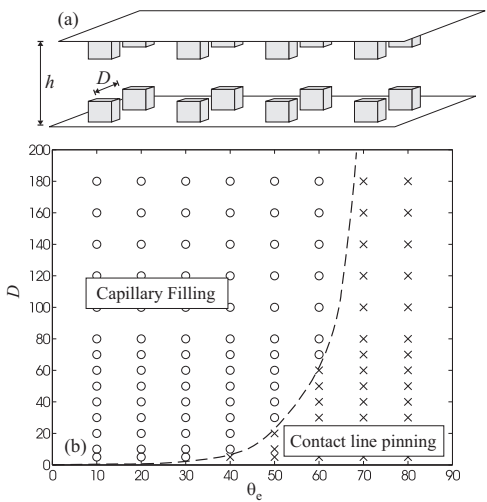}
\caption{Two dimensional surface patterning: (a) the surface geometry; (b) regions of 
contact line pinning and capillary flow as a function of post separation and equilibrium 
contact angle. The channel height is kept constant at $h=50$. The dashed line is a guide 
to the eye.}
\label{fig5}
\end{figure}
We now consider two-dimensional surface patterning. Both walls of the
capillary are patterned with $20 \times 20$ square posts of height
10. The posts face each other across the channel, which is of height $h=50$.
They are separated by spaces of length $D$ in the direction
perpendicular to the flow. This geometry is illustrated in Fig \ref{fig5}(a),
which shows a side view of the channel. For $D=0$, the case corresponding to ridges across the channel, the
Gibbs' criterion predicts that the interface will always remain pinned
between the trailing edges of a pair of opposing posts. For $D \neq 0$,
however, the motion can be driven by the portion of the interface that lies between
the posts.

Fig \ref{fig5}(b) shows the crossover between capillary filling and contact line 
pinning as a function of $D$ and $\theta_e$.
For $\theta_e \lesssim 30^{\mathrm{o}}$, a small gap in between the posts is 
enough to allow filling. As the interface moves through the gap and protrudes 
beyond the posts, it is able to depin, initially from the side of the posts. 

For $\theta_e \gtrsim 60^{\mathrm{o}}$, however, filling is strongly suppressed.
This is due to two complementary effects: the force driving the interface though 
the gaps becomes weaker and the pinning on the sides of the posts becomes
stronger. Note that the part of the interface that lies in between the posts 
can bow out further as $D$ is increased and it is possible that the leading 
interface could touch the next row of posts, thus providing an alternative mechanism to 
depin the interface from the edges of the posts. 

We have also varied the relative height of the posts with respect to the channel width 
and find that the contact line pinning condition depends only very weakly on the 
channel height. This is because the interface behaviour is primarily determined by
depinning from the side edges of the posts. It is also worth noting that we neglect fluctuations 
in our simulations.

\begin{figure}
\includegraphics[scale = 0.4]{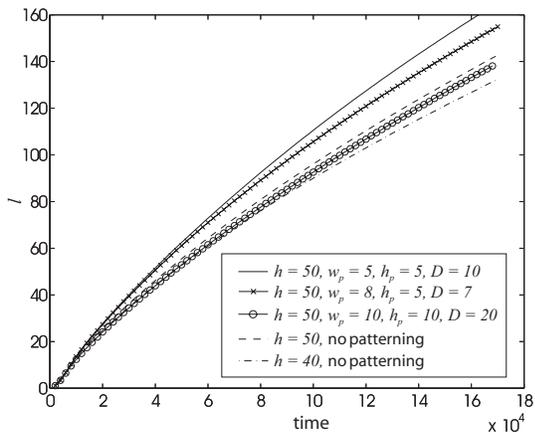}
\caption{The length of the column of $A$-liquid plotted against time for different  
surface patterning. The channels' intrinsic contact angle is 
$\theta_e = 30^{\mathrm{o}}$. Other parameters as in Fig. \ref{fig1}.}
\label{fig6}
\end{figure}
So far, we have considered surfaces patterned with ridges oriented perpendicular  
to the flow and square posts, cases where pinning by the posts 
suppresses, or even prevents capillary filling. We now identify a situation 
where patterning can be used to speed up capillary flow, ridges aligned
{\em along} the flow direction. In Fig. \ref{fig6}, we compare the lengths of
the $A$ column, plotted as a function of time, for a channel of height 
$h=50$ with infinitely long ridges of width ($w_p$), height ($h_p$) and separation ($D$)
(i) 5, 5, and 10, (ii) 8, 5, and 7, and (iii) 10, 10, and 20 lying along the channel. In all these 
cases the channels have the same surface roughness $r = 1 + 2h_p/(w_p+D) = 5/3$.
For comparison, we also show smooth channel simulations for $h=40$ and $h=50$. 
It is clear from Fig.  \ref{fig6}, that it is possible to increase the rate of 
capillary filling by patterning the channels (cases (i) and (ii)). This occurs 
because there is no contact angle hysteresis in this geometry and because the 
total surface area per unit length wetted by the $A$-liquid is increased over 
the flat channel value by the roughness factor $r$. However, from Fig. \ref{fig6}
it is apparent that the rate of filling is not just controlled by $r$. This is because 
surface roughness distorts the liquid flow from a parabolic profile and 
hence increases the dissipation. As shown in Fig. \ref{fig6}, the dissipation 
increases as the distance between two neighbouring ridges is reduced (case (ii))
and as the size of the ridges relative to the channel height is increased (case (iii)).
The filling speed is a competition between the increased energy gain of wetting a grooved
channel and the additional dissipation caused by the distortion of the flow.

Our results suggest strategies that could be used to design surface patterning on 
microfluidics channels. To speed up capillary filling, one should introduce roughness 
which is elongated in the direction of the flow. However, additional surface area will only 
speed up filling if pinning is avoided by, for example, avoiding sharp edges and making the post inclinations as gentle as 
possible. Furthermore, any surface patterning should neither be too dense nor too 
large to minimise dissipation. Random arrangement of the posts may also be 
used to avoid cooperative interface pinning effect. Finally, the intrinsic contact
angle of the materials $\theta_e \lesssim 60^{\mathrm{o}}$ for capillary filling down a
rough channel. These results are consistent with recent 
findings by Kohonen \cite{Kohonen}, where he found that the walls of the tracheids in 
the dry-habitat species are typically rougher than those in wet-habitat species, and with 
$\theta_e \sim 40^{\mathrm{o}}$. Moreover, the set of design principles we describe above 
may help explain why certain types of wall sculpturing, such as helical thickening 
\cite{Jeje} found on tree capillaries, are more common than others. 

We thank L. Biferale and S. Succi for useful discussions. HK acknowledges support from 
a Clarendon Bursary and the INFLUS project.

\end{document}